\documentclass[aps,prlonecolumn]{revtex4}
\usepackage{geometry}
\usepackage{mathptmx}
\usepackage{epsfig,amsopn}
\usepackage{amsmath,amssymb}
\usepackage{natbib}
\usepackage{xcolor}
\usepackage{braket}
\usepackage{float}
\usepackage{enumerate}
\usepackage{mathtools}
\usepackage[symbol]{footmisc}
\allowdisplaybreaks[1]
\usepackage{amsmath}
\usepackage{ragged2e}
\usepackage{amsmath}
\usepackage{mathptmx}
\usepackage{epsfig,amsopn}
\usepackage{graphicx}
\usepackage{amsmath,amssymb}
\usepackage{natbib,color}
\usepackage{braket}
\usepackage{float}
\usepackage{enumerate}
\usepackage{hyperref}
\allowdisplaybreaks[1]

\newtheorem{theorem}{Theorem}

\newtheorem{proposition}[theorem]{Proposition}

\begin{document}

\title{Poisson-process limit-laws yield Gumbel Max-Min and Min-Max}

\author{{\normalsize{}Iddo Eliazar$^1$}{\normalsize{}}}
\email{eliazar@post.tau.ac.il}

\author{{\normalsize{}Ralf Metzler$^2$}{\normalsize{}}}
\email{rmetzler@uni-potsdam.de}

\author{{\normalsize{}Shlomi Reuveni$^1$}{\normalsize{}}}
\email{shlomire@tauex.tau.ac.il}

\affiliation{\noindent \textit{
$^{1}$School of Chemistry, The Center for Physics and Chemistry of Living Systems, The Raymond and Beverly Sackler Center for Computational Molecular and Materials Science,\\ \& The Mark Ratner Institute for Single Molecule Chemistry, Tel Aviv University, Tel Aviv 6997801, Israel\\ $^{2}$University of Potsdam, Institute of Physics \& Astronomy, 14476
Potsdam, Germany.}}

\date{\today}

\begin{abstract}
\textquotedblleft A chain is only as strong as its weakest
link\textquotedblright\ says the proverb. But what about a collection of
statistically identical chains: How long till all chains fail? The answer to
this question is given by the Max-Min of a matrix whose $\left( i,j\right) $
entry is the failure time of link $j$ of chain $i$: take the minimum of each
row, and then the maximum of the rows' minima. The corresponding Min-Max is
obtained by taking the maximum of each column, and then the minimum of the
columns' maxima. The Min-Max applies to the storage of critical data.
Indeed, consider multiple backup copies of a set of critical data items, and
consider the $\left( i,j\right) $ matrix entry to be the time at which item $%
j$ on copy $i$ is lost; then, the Min-Max is the time at which the first
critical data item is lost. In this paper we address random matrices whose entries are independent and identically distributed random variables. We establish Poisson-process limit-laws for the row's minima and for the columns' maxima. Then, we further establish Gumbel limit-laws for the Max-Min and for the Min-Max. The limit-laws hold whenever the entries' distribution has a density, and the Gumbel limit-laws yield highly applicable approximation tools and design tools for large random matrices. 

\ 

\ 


\bigskip\ 

Brownian motion)
\end{abstract}
\maketitle

\newpage

\section{\label{1}Introduction}

\emph{Extreme Value Theory} (EVT) is a branch of probability theory that
focuses on extreme-value statistics such as maxima and minima \cite{Gal}-%
\cite{HF}. EVT has major applications in science and engineering \cite{Cas}-%
\cite{BGS}; examples range from insurance to finance, and from hydrology to
computer vision \cite{RT}-\cite{Sch}. At the core of EVT stands its
fundamental theorem, the Fisher-Tippett-Gnedenko theorem \cite{FT}-\cite{Gne}%
, which establishes the three extreme-value laws: \emph{Weibull} \cite{Wei1}-%
\cite{Wei2}, \emph{Frechet} \cite{Fre}, and \emph{Gumbel} \cite{Gum}.

The fundamental theorem of EVT applies to ensembles of independent and
identically distributed (IID) real-valued random variables, and is described
as follows \cite{BGT}. Consider an ensemble $\left\{ X_{1},\cdots
,X_{n}\right\} $ whose $n$ components are IID copies of a general
real-valued random variable $X$. Further consider the ensemble's maximum $%
M_{n}=\max \left\{ X_{1},\cdots ,X_{n}\right\} $, and an affine scaling of
this maximum: 
\begin{equation}
\tilde{M}_{n}=s_{n}\cdot \left( M_{n}-\delta _{n}\right) \text{ ,}
\label{101}
\end{equation}%
where $s_{n}$ is a positive scale parameter, and where $\delta _{n}$ is a
real location parameter. The fundamental theorem of EVT explores the
convergence in law (as $n\rightarrow \infty $) of the scaled maximum $\tilde{%
M}_{n}$ to a non-trivial limiting random variable $\mathcal{L}$.

Firstly, the fundamental theorem determines its admissible `inputs': the
classes of random variables $X$ that yield non-trivial limits $\mathcal{L}$.
Secondly, given an admissible input $X$, the fundamental theorem specifies
the adequate scale parameter $s_{n}$ and location parameter $\delta _{n}$.
Thirdly, as noted above, the fundamental theorem establishes that its
`outputs' are the three extreme-value laws: the statistics of the
non-trivial limits $\mathcal{L}$ are either Weibull, Frechet, or Gumbel. The 
\emph{domain of attraction} of each extreme-value law is the class of inputs 
$X$ yielding, respectively, each output law.

The fundamental theorem of EVT yields asymptotic approximations for the
maxima of large ensembles of IID real-valued random variables. Indeed,
consider the scaled maximum $\tilde{M}_{n}$ to converge in law (as $%
n\rightarrow \infty $) to a non-trivial limit $\mathcal{L}$. Then, for a
given large ensemble ($n\gg1$), the ensemble's maximum $M_{n}$ admits the
following \emph{extreme-value asymptotic approximation} in law: 
\begin{equation}
M_{n}\simeq \mathcal{L}_{\ast }:=\delta _{n}+\frac{1}{s_{n}}\cdot \mathcal{L}%
\text{ .}  \label{102}
\end{equation}

The extreme-value asymptotic approximation of Eq. (\ref{102}) has the
following meaning: the deterministic asymptotic approximation of the
ensemble's maximum $M_{n}$ is the location parameter $\delta _{n}$; the
magnitude of the random fluctuations about the deterministic asymptotic
approximation is $1/s_{n}$, the inverse of the scale parameter $s_{n}$; and
the statistics of the random fluctuations about the deterministic asymptotic
approximation are that of the limit $\mathcal{L}$ -- which is governed by
one of the three extreme-value laws.

The three extreme-value laws are \emph{universal} in the sense that they are
the \emph{only} non-trivial limiting statistics obtainable (as $n\rightarrow
\infty $) from the scaled maximum $\tilde{M}_{n}$. However, universality
holds neither for the corresponding domains of attraction, nor for the
corresponding scale parameter $s_{n}$ and location parameter $\delta _{n}$.
Indeed, each extreme-value law has a very specific and rather narrow domain
of attraction \cite{BGT}. Also, for any given admissible input $X$, the
scale parameter $s_{n}$ and location parameter $\delta _{n}$ are `custom
tailored' in a very precise manner \cite{BGT}.

In essence, the fundamental theorem of EVT considers a random-vector
setting: the maxima of what can be perceived as vector-structured ensembles
of IID real-valued random variables. This paper elevates from the
random-vector setting to the following \emph{random-matrix} setting: the 
\emph{Max-Min} and the \emph{Min-Max} of matrix-structured ensembles of IID
real-valued random variables. The Max-Min is obtained by taking the minimum
of each matrix-row, and then taking the maximum of the rows' minima. The
Min-Max is obtained by taking the maximum of each matrix-column, and then
taking the minimum of the columns' maxima.

The Max-Min and the Min-Max values of matrices emerge naturally in science
and engineering. Perhaps the best known example of the Max-Min and the
Min-Max comes from \emph{game theory} \cite{FuT}-\cite{MSZ}. Indeed,
consider a player that has a set of admissible strategies, and that faces a
set of viable scenarios. A payoff matrix determines the player's gains --
or, alternatively, losses -- for each strategy it applies and for each
scenario it encounters. The player's goal is to optimize with respect to the 
\emph{worst-case scenario}. Hence, in the case of gains, the player goes 
\emph{Max-Min}: calculate the minimal gain per each scenario, and then pick
the strategy that yields the largest minimal gain. And, in the case of
losses, the player goes \emph{Min-Max}: calculate the maximal loss per each
scenario, and then pick the strategy that yields the smallest maximal loss.
In the field of game theory the Max-Min and the Min-Max values appear also
in the context of game-searching procedures on trees \cite{Pea}-\cite{KDN}.

Architectural illustrations of the Max-Min and the Min-Max values come from 
\emph{reliability engineering} \cite{BP}-\cite{Fin}, where one is interested
in calculating the failure time (or the failure load) of a given system. Two
important system-architectures are, so called, \textquotedblleft
series-parallel\textquotedblright\ and \textquotedblleft
parallel-series\textquotedblright\ \cite{Kolo,Kolo1,Kolo2}. In the series-parallel
architecture a system is a parallel array of sub-systems, and each
sub-system is a serial array of components. In the parallel-series
architecture a system is a serial array of sub-systems, and each sub-system
is a parallel array of components. The Max-Min and the Min-Max values
correspond, respectively, to the failure times (or the failure loads) of
systems with series-parallel\ and with parallel-series architectures \cite{Kolo,Kolo1,Kolo2}.

There are several limit-law results -- counterparts of the fundamental theorem of EVT -- for the Max-Min and the Min-Max of random matrices (with IID entries). The pioneering mathematical results were presented by Chernoff and Teicher \cite{CT}, reliability-engineering results were presented by Kolowrocki \cite{Kolo}-\cite{Kolo2}, and relatively recent reliability-engineering results were presented by Reis and Castro \cite{RC}. All these limit-law results use affine scalings -- similar to that of Eq. (\ref{101}) -- for the Max-Min and the Min-Max. Also, all these limit-law results employ asymptotic couplings of the dimensions of the random matrices (as these dimensions are taken to infinity). 

Chernoff and Teicher established that the limit-laws for the Max-Min and the Min-Max are the three extreme-value laws \cite{CT}: Weibull, Frechet, and Gumbel. Kolowrocki investigated limit-laws for the Max-Min and the Min-Max in the context of systems with the aforementioned series-parallel and parallel-series architectures \cite{Kolo},\cite{Kolo1}-\cite{Kolo2}. Considering the Max-Min, and applying the fundamental theorem of EVT iteratively -- first to the minimum of each matrix-row, and then to the maximum of the rows' minima -- Reis and Castro established a Gumbel limit-law \cite{RC}; this limit-law applies to matrix entries that belong to sub-sets of the domains of attraction of the three extreme-value laws. 

For the results of \cite{CT}-\cite{RC} -- as in the case of the fundamental theorem of EVT -- universality holds neither with regard to the domains of attraction, nor with regard to the affine scalings. Also, for these results, universality does not hold with regard to the asymptotic couplings of the dimensions of the random matrices. Moreover, as the results of \cite{CT}-\cite{RC} involve very intricate mathematical conditions and schemes, their practical implementation is extremely challenging.

The limit-law results of \cite{CT}-\cite{RC} are derived via an `EVT machinery', i.e. methods similar to the Fisher-Tippett-Gnedenko theorem, together with other EVT results (e.g. \cite{BdH}). In this paper we take an altogether different approach: a `bedrock' Poisson-process method. Specifically, we dive down to the bedrock level of the rows' minima and the columns' maxima (of random matrices with IID entries), and establish \emph{Poisson-process limit-laws} for these minima and maxima. Then, elevating back from the bedrock level to the Max-Min and the Min-Max, we establish \emph{Gumbel limit-laws} for these values.   

The limit-laws presented here have the following key features. Firstly, their domain of attraction is vast: the limit-laws hold whenever the entries' distribution has a density. Secondly, they use affine scalings similar to that of Eq. (\ref{101}) with: a location parameter that is tunable (it can be set as we wish within the interior of the support of the IID entries); and a scale parameter that depends on the entries' distribution only up to a coefficient. Thirdly, their asymptotic couplings (of the dimensions of the random matrices) are geometric. Due to these features the practical implementation of the limit-laws presented here is easy and straightforward, and hence these results are highly applicable.

Figure 1 demonstrates the potency of the Gumbel limit-law for the Max-Min (see section \ref{3} for the details). This figure depicts numerical simulations of the Max-Min of random matrices whose IID entries are drawn from an assortment of distributions: Exponential, Gamma, Log-Normal, Inverse-Gauss, Uniform, Weibull, Beta, Pareto, and Normal. For all these distributions, the convergence of the simulations to the theoretical prediction of the Max-Min result is evident. The MATLAB code that was used in order to generate the simulations is detailed in the Appendix; this short code shows just how easy it is to apply, in practice, the novel Gumbel limit-laws presented here.

The reminder of this paper is organized as follows. Section \ref{2} presents the random-matrix setting, and the `bedrock' Poisson-process limit-law for the rows' minima. Then, section \ref{3} establishes the Gumbel limit-law for the Max-Min -- which is motivated by the following question: within a collection of IID chains, how long will the strongest chain hold? Section \ref{4} further establishes the counterpart Gumbel limit-law for the Min-Max -- which is based on a counterpart `bedrock' Poisson-process
limit-law for the columns' maxima, and which is motivated by the following question: using a collection of IID data-storage backup copies, how long can the data be stored reliably by the backup copies? Section \ref{5} describes the application of the Gumbel limit-laws as approximation tools and as design tools. An in-depth discussion of the limit-laws is held in section \ref{6}. Finally, section \ref{7} concludes, and the proofs of the key results stated along the paper are detailed in the Appendix.

\section{\label{2}Bedrock}

Consider a collection of $c$ \emph{chains}, labeled by the index $i=1,\cdots
,c$. Each chain comprises of $l$ \emph{links}, and all the $c\cdot l$ links
are IID copies of a generic link. In this paper we take a temporal
perspective and associate the \emph{failure time} of the generic link with a
real-valued random variable $T$. Namely, $T$ is the random time at which the generic link fails mechanically.

As the analysis to follow is probabilistic, we introduce relevant
statistical notation. Denote by $F\left( t\right) =\Pr \left( T\leq t\right) 
$ ($-\infty <t<\infty $) the \emph{distribution function} of the generic
failure time $T$, and by $\bar{F}\left( t\right) =\Pr \left( T>t\right) $ ($%
-\infty <t<\infty $) the corresponding \emph{survival function}. These
functions are coupled by $F\left( t\right) +\bar{F}\left( t\right) =1$ ($%
-\infty <t<\infty $). The \emph{density function} of the generic failure
time $T$ is given by $f\left( t\right) =F^{\prime }\left( t\right) =-\bar{F}%
^{\prime }\left( t\right) $ ($-\infty <t<\infty $). In particular, this notation covers the case of a positive-valued generic failure time $T$. We note that, alternative to the temporal perspective taken here, the random variable $T$ can manifest any other real-valued quantity of interest of the generic link, e.g. its mechanical strength (in which case $T$ is positive-valued). 

The following \emph{random matrix} underlies the collection of chains:%
\begin{equation}
\mathbf{T}=\left( 
\begin{array}{ccc}
T_{1,1} & \cdots & T_{1,l} \\ 
\vdots & \ddots & \vdots \\ 
T_{c,1} & \cdots & T_{c,l}%
\end{array}%
\right) \text{ .}  \label{201}
\end{equation}%
The dimensions of the random matrix $\mathbf{T}$ are $c\times l$, and its
entries are IID copies of the generic failure time $T$. The $i^{\text{th}}$
row of the random matrix $\mathbf{T}$ represents the $l$ links of chain $i$,
and the entries of this row manifest the respective failure times of the
links of chain $i$. Specifically, the entry $T_{i,j}$ is the failure time of
link $j$ of chain $i$.

\textquotedblleft A chain is only as strong as its weakest
link\textquotedblright\ says the proverb. So, chain $i$ fails as soon as one
of its links fails. Hence the chain's failure time is given by the \emph{%
minimum} of the failure times of its links:%
\begin{equation}
\wedge _{i}=\min \left\{ T_{i,1},\cdots ,T_{i,l}\right\}  \label{202}
\end{equation}%
($i=1,\cdots ,c$). Namely, the random variable $\wedge _{i}$ is the minimum
over the entries of the$\ i^{\text{th}}$ row of the random matrix $\mathbf{T}
$.

Now, consider an arbitrary reference time $t_{\ast }$ of the generic failure
time $T$, e.g. its median, its mean (in case the mean is finite), or its
mode (in case the density function $f\left( t\right) $ is unimodal). In
general, the reference time $t_{\ast }$ can be any real number that
satisfies two basic requirements: (i) $0<F\left( t_{\ast }\right) <1$, which
is equivalent to $0<\bar{F}\left( t_{\ast }\right) <1$; and (ii) $0<f\left(
t_{\ast }\right) <\infty $. These requirements are met by all the interior
points in the support of the input $T$.

With respect to the reference time $t_{\ast }$, we apply the following
affine scaling to the failure time of the $i^{\text{th}}$ chain:%
\begin{equation}
\tilde{\wedge}_{i}=l\cdot \left( \wedge _{i}-t_{\ast }\right)  \label{203}
\end{equation}%
($i=1,\cdots ,c$). Namely, in the affine scaling of Eq. (\ref{203}) the
chains' common length $l$ is the positive scale parameter, and the reference
time $t_{\ast }$ is the real location parameter.

Our goal is to analyze the limiting behavior of the chains' scaled
failure times in the case of a multitude of long chains: $c\rightarrow
\infty $ and $l\rightarrow \infty $. To that end we set our focus on the
ensemble of the chains' scaled failure times: $\left\{ \tilde{\wedge}%
_{1},\cdots ,\tilde{\wedge}_{c}\right\} $. Also, to that end we introduce
the following \emph{asymptotic geometric coupling} between the number $c$ of
the chains and the common length $l$ of the chains: $c\cdot \bar{F}\left(
t_{\ast }\right) ^{l}\simeq 1$. Specifically, the asymptotic geometric
coupling is given by the limit 
\begin{equation}
\lim_{c\rightarrow \infty ,l\rightarrow \infty }c\cdot \bar{F}\left( t_{\ast
}\right) ^{l}=1\text{ .}  \label{204}
\end{equation}

With the affine scaling of Eq. (\ref{203}), and the asymptotic geometric
coupling of Eq. (\ref{204}), we are now in position to state the following
Poisson-process limit-law result.

\begin{proposition}
\label{P1}The ensemble $\left\{ \tilde{\wedge}_{1},\cdots ,\tilde{\wedge}%
_{c}\right\} $ converges in law, in the limit of Eq. (\ref{204}), to a
limiting ensemble $\mathcal{P}$ that is a Poisson process over the real line
with the following intensity function: $\lambda \left( x\right) =\bar{%
\epsilon}\exp \left( -\bar{\epsilon}x\right) $ ($-\infty <$ $x<\infty $),
where $\bar{\epsilon}=f\left( t_{\ast }\right) /\bar{F}\left( t_{\ast
}\right) $.
\end{proposition}

See the Appendix for the proof of proposition \ref{P1}. Table 1 summarizes
proposition \ref{P1} and its underlying setting. We now elaborate on the
meaning of this proposition.

A \emph{Poisson process} is a countable collection of points that are
scattered randomly over its domain, according to certain Poisson-process
statistics that are determined by its \emph{intensity function} \cite{Kin}-%
\cite{Str}. Poisson processes are of key importance in probability theory,
and their applications range from insurance and finance \cite{EKM} to
queueing systems \cite{Wol}, and from fractals \cite{LT} to power-laws \cite%
{PWPL}.

In the case of the Poisson process $\mathcal{P}$ of proposition \ref{P1} the
domain is the real line ($-\infty <$ $x<\infty $), and the intensity
function is $\lambda \left( x\right) =\bar{\epsilon}\exp \left( -\bar{%
\epsilon}x\right) $. The points of the Poisson process $\mathcal{P}$ of
proposition \ref{P1} manifest, in the limit of Eq. (\ref{204}), the chains'
scaled failure times. The informal meaning of the intensity function $%
\lambda \left( x\right) $ is the following: the probability that the
infinitessimal interval $\left( x,x+dx\right) $ contains a point of the
Poisson process $\mathcal{P}$ is $\lambda \left( x\right) dx$, and this
probability is independent of the scattering of points outside the interval $%
\left( x,x+dx\right) $.

The exponent $\bar{\epsilon}=f\left( t_{\ast }\right) /\bar{F}\left( t_{\ast
}\right) $ of the intensity function $\lambda \left( x\right) $ manifests
the \emph{hazard rate} of the generic failure time $T$ at time $t_{\ast }$ 
\cite{BP}-\cite{Fin}: $\bar{\epsilon}$ is the likelihood that the generic
link will fail right after time $t_{\ast }$, conditioned on the information
that the generic link did not fail up to time $t_{\ast }$. Specifically,
this hazard rate is given by the following limit: 
\begin{equation}
\bar{\epsilon}=\lim_{\Delta \rightarrow 0}\frac{1}{\Delta }\Pr \left( T\leq
t_{\ast }+\Delta |T>t_{\ast }\right) \text{ .}  \label{205}
\end{equation}%
The hazard rate is a widely applied tool in reliability engineering and in
risk management \cite{BP}-\cite{Fin}.

\section{\label{3}Max-Min}

With proposition \ref{P1} at our disposal, we now set the
focus on the \emph{strongest chain}, i.e. the last chain standing. The failure time of the strongest chain is
given by the \emph{maximum} of the chains' failure times:%
\begin{equation}
\ \wedge _{\max }=\max \left\{ \wedge _{1},\cdots ,\wedge _{c}\right\} \text{
.}  \label{301}
\end{equation}%
Namely, the random variable $\wedge _{\max }$ is the \emph{Max-Min} over the
entries of the random matrix $\mathbf{T}$: for each and every row of the
matrix pick the minimal entry, and then pick the rows' largest minimal entry.

As with the chains' failure times, we apply the affine scaling of Eq. (\ref%
{203}) to the failure time of the strongest chain:%
\begin{equation}
\ \tilde{\wedge}_{\max }=l\cdot \left( \ \wedge _{\max }-t_{\ast }\right) 
\text{ ,}  \label{302}
\end{equation}%
where $t_{\ast }$ is the above reference time. Also, as with the ensemble $%
\left\{ \tilde{\wedge}_{1},\cdots ,\tilde{\wedge}_{c}\right\} $, we analyze
the limiting behavior of the random variable $\tilde{\wedge}_{\max }$ in the
case of a multitude of long chains: $c\rightarrow \infty $ and $l\rightarrow
\infty $.

Here and hereinafter $\mathcal{G}$ denotes a `standard' \emph{Gumbel} random
variable. Namely, $\mathcal{G}$ is a real-valued random variable whose
statistics are governed by the following `standard' Gumbel distribution
function: 
\begin{equation}
\Pr \left( \mathcal{G}\leq t\right) =\exp \left[ -\exp \left( -t\right) %
\right]  \label{303}
\end{equation}%
($-\infty <t<\infty $). We note that within the three extreme-value laws,
Gumbel is the only law whose range is the entire real line.

The three extreme-value laws have one-to-one correspondences with the
maximal points of specific Poisson processes \cite{GinEVT}. In particular,
the Gumbel extreme-value law has a one-to-one correspondence with the
maximal point of the Poisson process $\mathcal{P}$ of proposition \ref{P1}.
This connection leads to the following Gumbel limit-law result.

\begin{proposition}
\label{P2}The random variable $\tilde{\wedge}_{\max }$ converges in law, in
the limit of Eq. (\ref{204}), to a limiting random variable $\bar{\eta}\cdot 
\mathcal{G}$, where $\bar{\eta}=\bar{F}\left( t_{\ast }\right) /f\left(
t_{\ast }\right) $, and where $\mathcal{G}$ is the `standard' Gumbel random
variable of Eq. (\ref{303}).
\end{proposition}

See the Appendix for the proof of proposition \ref{P2}. Table 2 summarizes
proposition \ref{P2} and its underlying setting. In Figure 1 we use
numerical simulations to demonstrate Proposition \ref{P2}. To that end nine
different distributions of the generic failure time $T$ are considered:
Exponential, Gamma, Log-Normal, Inverse-Gauss, Uniform, Weibull, Beta,
Pareto, and Normal. In all nine cases, the convergence of the simulations to
the theoretical prediction of proposition \ref{P2} is evident. See the
Appendix for the MATLAB\ code that was used in order to generate the
numerical simulations.

Proposition \ref{P2} yields an asymptotic approximation for the Max-Min of
large random matrices with dimensions $c>l\gg1$. Indeed, consider the
matrix-dimensions ($c$ and $l$) and the reference time ($t_{\ast }$) to
satisfy the relation $c\cdot \bar{F}\left( t_{\ast }\right) ^{l}\simeq 1$.
Then, the Max-Min random variable $\wedge _{\max }$ admits the following 
\emph{Gumbel asymptotic approximation} in law:%
\begin{equation}
\ \wedge _{\max }\simeq \mathcal{G}_{\ast }:=t_{\ast }+\frac{\bar{\eta}}{l}%
\cdot \mathcal{G}\text{ ,}  \label{304}
\end{equation}%
where $\bar{\eta}$ and $\mathcal{G}$ are as in proposition \ref{P2}.

The Gumbel asymptotic approximation of Eq. (\ref{304}) has the following
meaning: the deterministic asymptotic approximation of the Max-Min $\wedge
_{\max }$ is the reference time $t_{\ast }$; the magnitude of the random
fluctuations about the deterministic asymptotic approximation is $\bar{\eta}%
/l$; and the statistics of the random fluctuations about the deterministic
asymptotic approximation are Gumbel. Table 3 summarizes the Gumbel
asymptotic approximation of Eq. (\ref{304}), and details the key statistical
features of this approximation.

\section{\label{4}Min-Max}

So far we addressed the \emph{Max-Min} of the random matrix $\mathbf{T}$:
pick the minimum of each row $\wedge _{i}=\min \left\{ T_{i,1},\cdots
,T_{i,l}\right\} $ ($i=1,\cdots ,c$), and then pick the maximum of these
minima $\wedge _{\max }=\max \left\{ \wedge _{1},\cdots ,\wedge _{c}\right\} 
$. Analogously, we can address the \emph{Min-Max} of the random matrix $%
\mathbf{T}$: pick the maximum of each column 
\begin{equation}
\vee _{j}=\max \left\{ T_{1,j},\cdots ,T_{c,j}\right\}  \label{401}
\end{equation}%
($j=1,\cdots ,l$), and then pick the minimum of these maxima 
\begin{equation}
\vee _{\min }=\min \left\{ \vee _{1},\cdots ,\vee _{l}\right\} \text{ .}
\label{402}
\end{equation}

To illustrate the Min-Max $\vee _{\min }$ consider the collection of the
aforementioned $c$ chains to be copies of a given DNA strand. The chains' $l$
links represent $l$ sites along the DNA strand, where each of these sites
codes a critical information item. The links' generic failure time $T$
manifests the time at which the information coded by a specific DNA\ site is
damaged; namely, the matrix entry $T_{i,j}$ is the time at which the $j^{%
\text{th}}$ information item on the $i^{\text{th}}$ DNA copy is damaged. The 
$j^{\text{th}}$ information item is lost once all its $c$ copies are
damaged, and hence the failure time of the $j^{\text{th}}$ information item
is given by Eq. (\ref{401}). As all the $l$ information items are critical,
a system-failure occurs once any of the $l$ information items is lost.
Hence, the time of the system-failure is given by the \emph{Min-Max} of Eq. (%
\ref{402}).

More generally, the Min-Max $\vee _{\min }$ applies to a setting in which $l$
critical information items are stored on $c$ different backup copies, where: 
$j=1,\cdots ,l$ is the index of the information items; $i=1,\cdots ,c$ is
the index of the copies; and $T_{i,j}$ is the time at which the $j^{\text{th}%
}$ information item on the $i^{\text{th}}$ backup copy is damaged. The above
`DNA model' was for the sake of illustration -- following the `chains model'
of section \ref{2}, which we used in order to illustrate the Max-Min.

The analysis presented above was with regard to the Max-Min. Analogous
analysis holds with regard to the Min-Max. Indeed, consider the above
reference time $t_{\ast }$, and apply the following affine scaling to the
failure time of the $j^{\text{th}}$ information item: 
\begin{equation}
\tilde{\vee}_{j}=c\cdot \left( \vee _{j}-t_{\ast }\right)  \label{403}
\end{equation}%
($j=1,\cdots ,l$). Namely, in the affine scaling of Eq. (\ref{403}) the
number $c$ of the copies is the positive scale parameter, and the reference
time $t_{\ast }$ is the real location parameter.

Also, introduce an \emph{asymptotic geometric coupling} between the number $%
l $ of the information items and the number $c$ of the copies: $l\cdot
F\left( t_{\ast }\right) ^{c}\simeq 1$. Specifically, the asymptotic
geometric coupling is given by the limit 
\begin{equation}
\lim_{l\rightarrow \infty ,c\rightarrow \infty }l\cdot F\left( t_{\ast
}\right) ^{c}=1\text{ .}  \label{404}
\end{equation}

With the affine scaling of Eq. (\ref{203}), and the asymptotic geometric
coupling of Eq. (\ref{204}), we are now in position to state the following
counterpart of proposition \ref{P1}.

\begin{proposition}
\label{P3}The ensemble $\left\{ \tilde{\vee}_{1},\cdots ,\tilde{\vee}%
_{l}\right\} $ converges in law, in the limit of Eq. (\ref{404}), to a
limiting ensemble $\mathcal{P}$ that is a Poisson process over the real line
with the following intensity function: $\lambda \left( x\right) =\epsilon
\exp \left( \epsilon x\right) $ ($-\infty <$ $x<\infty $), where $\epsilon
=f\left( t_{\ast }\right) /F\left( t_{\ast }\right) $.
\end{proposition}

See the Appendix for the proof of proposition \ref{P3}. Table 1 summarizes
proposition \ref{P3} and its underlying setting. The notion of Poisson
processes was described right after proposition \ref{P1}. The exponential
intensity function $\lambda \left( x\right) =\epsilon \exp \left( \epsilon
x\right) $ of proposition \ref{P3}, and the Poisson process $\mathcal{P}$
that this intensity characterizes, are most intimately related to the notion
of \emph{accelerating change} \cite{AccCha}; readers interested in a
detailed analysis of the (rich) statistical structure of this Poisson
process are referred to \cite{AccCha}. The exponent $\epsilon =f\left(
t_{\ast }\right) /F\left( t_{\ast }\right) $ has the following limit
interpretation: 
\begin{equation}
\epsilon =\lim_{\Delta \rightarrow 0}\frac{1}{\Delta }\Pr (T>t_{\ast
}-\Delta |T\leq t_{\ast })\text{ ,}  \label{407}
\end{equation}%
which is a time-reversal analogue of the hazard rate of Eq. (\ref{205}).

Continuing on from proposition \ref{P3}, and considering the above reference
time $t_{\ast }$, we apply the affine scaling of Eq. (\ref{403}) to the time
of the system-failure: 
\begin{equation}
\tilde{\vee}_{\min }=c\cdot \left( \vee _{\min }-t_{\ast }\right) \text{ .}
\label{405}
\end{equation}%
Then, as proposition \ref{P1} led to proposition \ref{P2}, proposition \ref%
{P3} leads to the following Gumbel limit-law result -- which is the Min-Max
counterpart of proposition \ref{P2}.

\begin{proposition}
\label{P4}The random variable $\tilde{\vee}_{\min }$ converges in law, in
the limit of Eq. (\ref{404}), to a limiting random variable $-\eta \cdot 
\mathcal{G}$, where $\eta =F\left( t_{\ast }\right) /f\left( t_{\ast
}\right) $, and where $\mathcal{G}$ is the `standard' Gumbel random variable
of Eq. (\ref{303}).
\end{proposition}

See the Appendix for the proof of proposition \ref{P4}. Table 2 summarizes
proposition \ref{P4} and its underlying setting. Proposition \ref{P4} yields
an asymptotic approximation for the Min-Max of large random matrices with
dimensions $l>c\gg1$. Indeed, consider the matrix-dimensions ($l$ and $c$)
and the reference time ($t_{\ast }$) to satisfy the relation $l\cdot F\left(
t_{\ast }\right) ^{c}\simeq 1$. Then, the Min-Max random variable $\vee
_{\min }$ admits the following \emph{Gumbel asymptotic approximation} in law:%
\begin{equation}
\vee _{\min }\simeq \mathcal{G}_{\ast }:=t_{\ast }-\frac{\eta }{c}\cdot 
\mathcal{G}\text{ ,}  \label{406}
\end{equation}%
where $\eta $ and $\mathcal{G}$ are as in proposition \ref{P4}.

The Gumbel asymptotic approximation of Eq. (\ref{406}) is the Min-Max
counterpart of the Max-Min Gumbel asymptotic approximation of Eq. (\ref{304}%
). Specifically: the deterministic asymptotic approximation of the Min-Max $%
\vee _{\min }$ is the reference time $t_{\ast }$; the magnitude of the
random fluctuations about the deterministic asymptotic approximation is $%
\eta /c$; and the statistics of the random fluctuations about the
deterministic asymptotic approximation are Gumbel. Table 3 summarizes the
Gumbel asymptotic approximation of Eq. (\ref{406}), and details the key
statistical features of this approximation.

\section{\label{5}Application}

The Gumbel asymptotic approximations of Eq. (\ref{304}) and of Eq. (\ref{406}%
) can be applied in two modalities: as \emph{approximation tools} and as 
\emph{design tools }for the Max-Min and the Min-Max, respectively. Both
applications are based on the fact that -- for Eqs. (\ref{304}) and (\ref%
{406}) to hold -- it is required that the matrix-dimensions ($c$ and $l$)
and the reference time ($t_{\ast }$) be properly coupled. In this section we
describe and demonstrate these applications.

We start with the Max-Min, and its Gumbel asymptotic approximation of Eq. (%
\ref{304}). This approximation requires the following coupling between the
matrix-dimensions and the reference time: $c\cdot \bar{F}\left( t_{\ast
}\right) ^{l}\simeq 1$, where $c>l\gg1$. Consequently, if the
matrix-dimensions are given ($c>l\gg1$) then the approximation of Eq. (\ref%
{304}) holds with the following \emph{implied reference time}:%
\begin{equation}
t_{\ast }=\bar{F}^{-1}\left[ \left( \frac{1}{c}\right) ^{1/l}\right] \text{ .%
}  \label{501}
\end{equation}%
For example, if $c=2^{l}$ then the implied reference time is the \emph{median%
} of the generic failure time $T$. This application is an \emph{%
approximation tool}: given the random matrix $\mathbf{T}$, Eq. (\ref{304})
with the implied reference time of Eq. (\ref{501}) approximates the Max-Min
of the matrix.

To demonstrate the \emph{design-tool} application of the Gumbel asymptotic
approximation of Eq. (\ref{304}), consider a system with a \textquotedblleft
series-parallel\textquotedblright\ architecture: the system is a parallel
array of $c$ sub-systems (labeled $i=1,\cdots ,c$), and each sub-system is a
serial array of $l$ components (labeled $j=1,\cdots ,l$). In terms of the
random matrix $\mathbf{T}$ of Eq. (\ref{201}), the failure time of component 
$j$ in sub-system $i$ is $T_{i,j}$. The series-parallel architecture implies
that the system's failure time is the Max-Min $\wedge _{\max }$. Now, assume
that our goal is to design a system whose failure time has the following
properties: its deterministic approximation is $t_{\ast }$, and the
magnitude of its random fluctuations about its deterministic approximation
is $\bar{m}$ -- where $t_{\ast }$ and $\bar{m}$ are specified target values.
Then, to meet the goal, the dimensions of the system should be designed as
follows: 
\begin{equation}
l\simeq \frac{1}{\bar{m}}\frac{\bar{F}\left( t_{\ast }\right) }{f\left(
t_{\ast }\right) }\text{ \ \& \ }c\simeq \frac{1}{\bar{F}\left( t_{\ast
}\right) ^{l}}\text{ .}  \label{502}
\end{equation}

Let's turn now to the Min-Max, and its Gumbel asymptotic approximation of
Eq. (\ref{406}). This approximation requires the following coupling between
the matrix-dimensions and the reference time: $l\cdot F\left( t_{\ast
}\right) ^{c}\simeq 1$, where $l>c\gg1$. Consequently, if the
matrix-dimensions are given ($l>c\gg1$) then the approximation of Eq. (\ref%
{406}) holds with the following \emph{implied reference time}:%
\begin{equation}
t_{\ast }=F^{-1}\left[ \left( \frac{1}{l}\right) ^{1/c}\right] \text{ .}
\label{503}
\end{equation}%
For example, if $l=2^{c}$ then the implied reference time is the \emph{median%
} of the generic failure time $T$. This application is an \emph{%
approximation tool}: given the random matrix $\mathbf{T}$, Eq. (\ref{406})
with the implied reference time of Eq. (\ref{503}) approximates the Min-Max
of the matrix.

To demonstrate the \emph{design-tool} application of the Gumbel asymptotic
approximation of Eq. (\ref{406}), consider a system with a \textquotedblleft
parallel-series\textquotedblright\ architecture: the system is a serial
array of $l$ sub-systems (labeled $j=1,\cdots ,l$), and each sub-system is a
parallel array of $c$ components (labeled $i=1,\cdots ,c$). In terms of the
random matrix $\mathbf{T}$ of Eq. (\ref{201}), the failure time of component 
$i$ in sub-system $j$ is $T_{i,j}$. The parallel-series architecture implies
that the system's failure time is the Min-Max $\vee _{\min }$. Now, assume
that our goal is to design a system whose failure time has the following
properties: its deterministic approximation is $t_{\ast }$, and the
magnitude of its random fluctuations about its deterministic approximation
is $m$ -- where $t_{\ast }$ and $m$ are specified target values. Then, to
meet the goal, the dimensions of the system should be designed as follows:%
\begin{equation}
c\simeq \frac{1}{m}\frac{F\left( t_{\ast }\right) }{f\left( t_{\ast }\right) 
}\text{ \ \& \ }l\simeq \frac{1}{F\left( t_{\ast }\right) ^{c}}\text{ .}
\label{504}
\end{equation}

Eq. (\ref{501}) and Eq. (\ref{503}) are explicit formulae facilitating the
approximation of the Max-Min and Min-Max of large random matrices. Eq. (\ref%
{502}) and Eq. (\ref{504}) are explicit formulae facilitating the design of
systems with, respectively, \textquotedblleft
series-parallel\textquotedblright\ and \textquotedblleft
parallel-series\textquotedblright\ architectures. The practical
implementation of these formulae is easy and straightforward.

\section{\label{6}Discussion}

We opened this paper with the fundamental theorem of EVT, and with a short discussion of the extreme-value asymptotic approximations emerging from this theorem. We now continue with this discussion, and expand it to include the Gumbel asymptotic approximations of Eqs. (\ref{304}) and (\ref{406}), as well as the asymptotic approximation emanating from the Central Limit Theorem (CLT) of probability theory \cite{Fel1}-\cite{Fel2}. To that end we begin with a succinct review of the CLT. 

As in the case of the fundamental theorem of EVT, the CLT applies to
ensembles of IID real-valued random variables: $\left\{ X_{1},\cdots
,X_{n}\right\} $ where the ensemble's $n$ components are IID copies of a
general real-valued random variable $X$. The input $X$ is assumed to have a
finite (positive) standard deviation $\sigma $, and hence also a finite
(real) mean $\mu $. We consider the ensemble's average $A_{n}=\left(
X_{1}+\cdots +X_{n}\right) /n$, and further consider the following affine
scaling of this average: 
\begin{equation}
\tilde{A}_{n}=\frac{1}{\sigma }\sqrt{n}\cdot \left( A_{n}-\mu \right) \text{
.}  \label{611}
\end{equation}%
Eq. (\ref{611}) is the CLT counterparts of Eq. (\ref{101}) -- with the term $%
\sqrt{n}/\sigma $ assuming the role of the positive scale parameter ($s_{n}$
in Eq. (\ref{101})), and with the mean $\mu $ assuming the role of the real
location parameter ($\delta _{n}$ in Eq. (\ref{101})).

The CLT asserts that the scaled average $\tilde{A}_{n}$ convergence in law
(as $n\rightarrow \infty $) to a limiting random variable $\mathcal{N}$ that
is `standard' Normal; i.e. the statistics of the limit $\mathcal{N}$ are
Normal (Gauss) with zero mean and with unit variance. Consequently, for a
given large ensemble ($n\gg1$), the ensemble's average $A_{n}$ admits the
following \emph{Normal asymptotic approximation} in law:%
\begin{equation}
A_{n}\simeq \mathcal{N}_{\ast }:=\mu +\frac{\sigma }{\sqrt{n}}\cdot \mathcal{%
N}\text{ .}  \label{612}
\end{equation}%
The Normal asymptotic approximation of Eq. (\ref{612}) has the following
meaning: the deterministic asymptotic approximation of the ensemble's
average $A_{n}$ is the mean $\mu $; the magnitude of the random fluctuations
about the deterministic asymptotic approximation is $\sigma /\sqrt{n}$; and
the statistics of the random fluctuations about the deterministic asymptotic
approximation are Normal.

\bigskip

It is illuminating to compare the extreme-value asymptotic approximation of
Eq. (\ref{102}), the Normal asymptotic approximation of Eq. (\ref{612}), and
the Gumbel asymptotic approximations of Eqs. (\ref{304}) and (\ref{406}). Such a comparison will highlight the analogies and the differences between these asymptotic approximations -- as we shall now see.

The extreme-value asymptotic approximation of Eq. (\ref{102}) has the
following key features. (I) The domains of attraction are characterized by 
\emph{narrow tail conditions}: regular-variation conditions for the Weibull
and Frechet extreme-value laws, and a complicated condition for the Gumbel
extreme-value law (see theorems 8.13.2 - 8.13.4 in \cite{BGT}, and \cite{BdH}%
). (II) The deterministic asymptotic approximation $\delta _{n}$ is highly
dependent on the input $X$. (III) The fluctuations' magnitude $1/s_{n}$ is
highly dependent on the input $X$. (IV) The limit $\mathcal{L}$ is either Weibull, Frechet, or Gumbel. (V) The information required in order to apply this asymptotic approximation is infinite-dimensional: the input's distribution function.

The Normal asymptotic approximation of Eq. (\ref{612}) has the following key
features. (I) The domain of attraction is characterized by a \emph{wide moment condition}: inputs $X$ with a finite variance. (II) The deterministic
asymptotic approximation $\mu $ is the input's mean. (III) The fluctuations'
magnitude $\sigma /\sqrt{n}$ depends on the input $X$ only via the
coefficient $\sigma $ (which is the input's standard deviation); hence the
asymptotic order $O\left( 1/\sqrt{n}\right) $ of the fluctuations' magnitude is independent of the input $X$. (IV) The limit $\mathcal{N}$ is `standard' Normal. (V) The information required in order to apply this asymptotic approximation is two-dimensional: the input's mean and standard deviation.

The Gumbel asymptotic approximations of Eqs. (\ref{304}) and (\ref{406}) -- for a preset reference time $t_{\ast }$ -- have the following key features. (I) The domain of attraction is characterized by a \emph{wide smoothness condition}: inputs $T$ with a
density function. (II)\ The deterministic asymptotic approximation $t_{\ast
} $ is the preset reference time. (III) The
fluctuations' magnitudes $\bar{\eta}/l$ and $\eta /c$ depend on the input $T$
only via the coefficients $\bar{\eta}$ and $\eta $, respectively; hence the
asymptotic orders $O\left( 1/l\right) $ and $O\left( 1/c\right) $ of the
fluctuations magnitudes are independent of the input $T$. (IV) The limit $\mathcal{G}$ is `standard' Gumbel. (V) The information required in order to apply these asymptotic approximations is two-dimensional: the value of the input's distribution function and density function at the reference time $t_{\ast }$. 

On the one hand, the key features of the Gumbel asymptotic approximations of Eqs. (\ref{304}) and (\ref{406}) are quite different from those of the extreme-value asymptotic approximation of Eq. (\ref{102}). On the other, the key features of these Gumbel asymptotic approximations are markedly similar to those of the Normal asymptotic approximation of Eq. (\ref{612}). Thus, the Gumbel asymptotic approximations presented here are `as universal' as the Normal asymptotic approximation; the similarities between these approximations are summarized in Table 4.

\bigskip

As its name suggests, a cornerstone of the Central Limit Theorem (CLT) is
its \emph{centrality}. In terms of the Normal asymptotic approximation of
Eq. (\ref{612}), centrality is manifested as follows: the ensemble's average 
$A_{n}$ is approximated about the `center point' of the input $X$ -- its
mean $\mu $. In effect, the CLT `magnifies' the statistical behavior of the
ensemble's average $A_{n}$ about the `center point' $\mu $.

The fundamental theorem of EVT is diametric to the CLT. Indeed, denote by $%
x^{\ast }$ the upper bound of the support of the input $X$; this upper bound
can be either finite ($x^{\ast }<\infty $) or infinite ($x^{\ast }=\infty $%
). Specifically, in the Weibull case $x^{\ast }$ it is finite, in the
Frechet case $x^{\ast }$ it is infinite, and in the Gumbel case $x^{\ast }$
it is either (see theorems 8.13.2 - 8.13.4 in \cite{BGT}, and \cite{BdH}).
In effect, the fundamental theorem of EVT `magnifies' the statistical
behavior of the ensemble's maximum $M_{n}$ about the upper bound $x^{\ast }$.

Thus, on the one hand, the Normal asymptotic approximation of Eq. (\ref{612}%
) `anchors' at the mean $\mu $ -- which is an \emph{interior point} of the
support of the input $X$. And, on the other hand, the extreme-value
asymptotic approximation of Eq. (\ref{102}) `anchors' at the upper bound $%
x^{\ast }$ -- which is a \emph{boundary point} of the support of the input $X
$. So, also from an `anchoring perspective': the Gumbel asymptotic approximations of Eqs. (\ref{304}) and (\ref{406}) are different from the extreme-value asymptotic approximation of Eq. (\ref{102}), and are similar to the Normal asymptotic approximation of Eq. (\ref{612}). Indeed, these Gumbel asymptotic approximations `anchor' at the reference time $t_{\ast}$ -- which is an \emph{interior point} of the support of the input $T$.

Notably, in the design-tool modality, the Gumbel asymptotic approximations of Eqs. (\ref{304}) and (\ref%
{406}) offer a feature that even the CLT does not offer: \emph{tunability}.
The `center point' at which the Normal asymptotic approximation of Eq. (\ref%
{612}) `anchors' is the mean $\mu $ -- and this anchoring point is \emph{%
fixed}. The `center point' at which the Gumbel asymptotic approximations of
Eqs. (\ref{304}) and (\ref{406}) `anchor' is the reference time $t_{\ast }$
-- and this anchoring point is \emph{tunable}. Namely, propositions \ref{P1}-\ref{P4}
allow us to set the reference time $t_{\ast }$ as we wish within the support
of the input $T$.

\bigskip

Perhaps the most straightforward approach to tackle the Max-Min and the Min-Max of random matrices is to apply the fundamental theorem of EVT iteratively. Reis and Castro did precisely so for the Max-Min \cite{RC}: they applied the fundamental theorem first to the minimum of each and every row of the random matrix $\mathbf{T}$ (of Eq. (\ref{201})), and then to the maximum of the rows' minima. Interestingly, the results of Reis and Castro and our results both yield Gumbel limit-laws. Nonetheless, these seemingly identical limit-law results are profoundly different. ``God is in the details'' -- or in the features -- as we shall now elucidate.

Consider the iterative EVT approach. The first iteration of the fundamental theorem implicitly confines the input $T$ to one of the theorem's narrow domains of attraction (Weibull, Frechet, Gumbel); moreover, as noted above, this iteration `anchors' at the the upper bound of the support of the input $T$. To apply the second iteration one has to impose further conditions, as well as to introduce an asymptotic coupling between the dimensions of the random matrix $\mathbf{T}$. Consequently, the iterative EVT approach comes with an expensive `intricacy price tag'. Specifically, for the limit-law of \cite{RC} the following are highly dependent on the input $T$, and are also highly elaborate: the Max-Min domain of attraction, scaling scheme, and asymptotic coupling. Matters are as intricate also in the Max-Min and Min-Max results of \cite{CT}-\cite{Kolo2} (which are derived via `EVT machineries').

Here, rather than mimicking the fundamental theorem of EVT, we mimicked the CLT. Firstly, we set a vast domain of attraction: inputs $T$ with a
density function. Secondly, we devised particular asymptotic couplings and affine scalings: Eqs. (\ref{204}) and (\ref{302}) for the Max-Min, and Eqs. (\ref{404}) and (\ref{405}) for the Min-Max. Thirdly, we showed that these particular asymptotic couplings and affine scalings always yield the Gumbel limit-laws of propositions \ref{P2} and \ref{P4}; i.e. they do so for all inputs $T$ that belong to the vast domain of attraction. These novel Gumbel limit-laws were achieved via a Poisson-process approach: the `bedrock' Poisson-process limit-laws of propositions \ref{P1} and \ref{P3}. This approach enabled us to circumvent the use of the fundamental theorem of EVT.

The Gumbel limit-laws of propositions \ref{P2} and \ref{P4} are truly workable tools for the Max-Min and the Min-Max of random matrices with IID entries. In turn, so are the Gumbel asymptotic approximations of Eqs. (\ref{304}) and (\ref{406}). A short MATLAB code given in the Appendix shows just how easy it is to apply these tools in prctice. 

\section{\label{7}Conclusion}

This paper explored the Max-Min value $\wedge _{\max }$ and the Min-Max
value $\vee _{\min }$ of a random matrix $\mathbf{T}$ with: $c$ rows, $l$
columns, and entries that are IID real-valued random variables. This IID
setting is common to random-matrix theory, to the fundamental theorem of
Extreme Value Theory, and to the Central Limit Theorem. The Max-Min and the Min-Max values of
matrices emerge naturally in science and engineering, e.g. in game theory
and in reliability engineering. We motivated the Max-Min value $\wedge
_{\max }$ by the following question: within a collection of $c$ IID chains,
each with $l$ links, how long will the strongest chain hold? And, we
motivated the Min-Max value $\vee _{\min }$ by the following question: how
long can $l$ critical information items be stored reliably on $c$ IID backup
copies?

We showed that if the number of rows $c$ and the number of columns $l$ are
large, and are coupled geometrically, then: the Max-Min value $\wedge _{\max
}$ and the Min-Max value $\vee _{\min }$ admit, respectively, the Gumbel asymptotic approximations of Eq. (\ref{304}) and of
Eq. (\ref{406}) (in law). These Gumbel asymptotic approximations are
similar, in form, to the Normal asymptotic approximation that follows from
the Central Limit Theorem. Moreover, in their design-tool modality, the Gumbel asymptotic approximations
display a special feature: their deterministic part -- the reference time $%
t_{\ast }$ -- is tunable. Hence, these Gumbel asymptotic approximations can
be used, via Eqs. (\ref{502}) and (\ref{504}), to design the Max-Min and Min-Max values.

The Gumbel asymptotic approximations are founded on the Gumbel limit-laws of
propositions \ref{P2} and \ref{P4}. In turn, the Gumbel limit-laws are
founded on the `bedrock' Poisson-process limit-laws of propositions \ref{P1}
and \ref{P3}. These four novel limit-laws have a vast domain of attraction, have simple affine scalings, and use geometric asymptotic couplings (of $c$ and $l$). 
With their generality, their CLT-like structure, their straightforward practical implementation, and their many potential
applications -- the results established and presented in this paper are
expected to serve diverse audiences in science and engineering.

\newpage 

\textbf{Acknowledgments}. R.M. acknowledges Deutsche Forschungsgemeinschaft
for funding (ME 1535/7-1) and support from the Foundation for Polish Science
within an Alexander von Humboldt Polish Honorary Research Fellowship. S.R.
gratefully acknowledges support from the Azrieli Foundation and the Sackler
Center for Computational Molecular and Materials Science.

\section{\label{A}Appendix}

\subsection{\label{A0}A general Poisson-process limit-law result}

In this subsection we establish a general Poisson-process limit-law result.
The setting of the general result is as follows. Consider $X_{1},\cdots
,X_{n}$ to be $n$ IID copies of a generic random variable $X$. The random
variable $X$ is real-valued, and its density function is given by%
\begin{equation}
f_{\theta }\left( x\right) =\kappa _{\theta }\cdot g_{\theta }\left( x\right)
\label{A10}
\end{equation}%
($-\infty <x<\infty $), where: $\theta $ is a positive parameter; $\kappa
_{\theta }$ is a positive constant; $g_{\theta }\left( x\right) $ is a
non-negative function.

Consider the joint limits $n\rightarrow \infty $ and $\theta \rightarrow
\infty $. We assume that the parameter $n$ and the constant $\kappa _{\theta
}$ admit the following asymptotic coupling:%
\begin{equation}
\lim_{n\rightarrow \infty ,\theta \rightarrow \infty }n\cdot \kappa _{\theta
}=\kappa \text{ ,}  \label{A11}
\end{equation}%
where $\kappa $ is a positive limit value. Also, we assume that 
\begin{equation}
\lim_{\theta \rightarrow \infty }g_{\theta }\left( x\right) =g\left( x\right)
\label{A12}
\end{equation}%
($-\infty <x<\infty $), where $g\left( x\right) $ is a non-negative limit
function.

Now, let's analyze the asymptotic statistical behavior of the ensemble $%
\left\{ X_{1},\cdots ,X_{n}\right\} $ in the joint limits $n\rightarrow
\infty $ and $\theta \rightarrow \infty $. To that end we take a real-valued
`test function' $\phi \left( x\right) $ ($-\infty <x<\infty $), and compute
the \emph{characteristic functional} of the ensemble $\left\{ X_{1},\cdots
,X_{n}\right\} $ with respect to this test function:%
\begin{equation}
\left. 
\begin{array}{l}
\mathbf{E}\left[ \phi \left( X_{1}\right) \cdots \phi \left( X_{n}\right) %
\right] \\ 
\text{ } \\ 
=\mathbf{E}\left[ \phi \left( X\right) \right] ^{n}=\left\{ \int_{-\infty
}^{\infty }\phi \left( x\right) f_{\theta }\left( x\right) dx\right\} ^{n}
\\ 
\text{ } \\ 
=\left\{ 1-\int_{-\infty }^{\infty }\left[ 1-\phi \left( x\right) \right]
f_{\theta }\left( x\right) dx\right\} ^{n} \\ 
\text{ } \\ 
=\left\{ 1-\frac{1}{n}\int_{-\infty }^{\infty }\left[ 1-\phi \left( x\right) %
\right] \left[ \left( n\kappa _{\theta }\right) \cdot g_{\theta }\left(
x\right) \right] dx\right\} ^{n}%
\end{array}%
\right.  \label{A14}
\end{equation}%
(in Eq. (\ref{A14}) we used the IID structure of the ensemble $\left\{
X_{1},\cdots ,X_{n}\right\} $, and Eq. (\ref{A10})). Applying the limits of
Eqs. (\ref{A11})-(\ref{A12}), Eq. (\ref{A14}) implies that:%
\begin{equation}
\lim_{n\rightarrow \infty ,\theta \rightarrow \infty }\mathbf{E}\left[ \phi
\left( X_{1}\right) \cdots \phi \left( X_{n}\right) \right] =\exp \left\{
-\int_{-\infty }^{\infty }\left[ 1-\phi \left( x\right) \right] \left[
\kappa \cdot g\left( x\right) \right] dx\right\} \text{ .}  \label{A15}
\end{equation}

The \emph{characteristic functional} of a Poisson process $\mathcal{P}$ over
the real line, with intensity function $\lambda \left( x\right) $ ($-\infty
<x<\infty $), is given by \cite{Kin}:%
\begin{equation}
\mathbf{E}\left[ \prod_{x\in \mathcal{P}}\phi \left( x\right) \right] =\exp
\left\{ -\int_{-\infty }^{\infty }\left[ 1-\phi \left( x\right) \right]
\lambda \left( x\right) dx\right\} \text{ ,}  \label{A16}
\end{equation}%
where $\phi \left( x\right) $ ($-\infty <x<\infty $) is a real-valued `test
function'. We emphasize that the characteristic functional of Eq. (\ref{A16}) is indeed characteristic \cite{Kin}: if $\mathcal{P}$ is collection of real points that satisfies Eq. (\ref{A16}) -- then $\mathcal{P}$ is a Poisson process over the real line, with intensity function $\lambda \left( x\right) $ ($-\infty<x<\infty $). Hence, combined together, Eqs. (\ref{A15}) and (\ref{A16}) yield
the following general result:

\begin{proposition}
\label{P}The ensemble $\left\{ X_{1},\cdots ,X_{n}\right\} $\ converges in
law, in the joint limits $n\rightarrow \infty $ and $\theta \rightarrow
\infty $, to a Poisson process $\mathcal{P}$ over the real line with
intensity function $\lambda \left( x\right) =\kappa \cdot g\left( x\right) $
($-\infty <x<\infty $).
\end{proposition}

\subsection{\label{A1}Proof of proposition \protect\ref{P1}}

Eq. (\ref{202}) implies that%
\begin{equation}
\left. 
\begin{array}{l}
\Pr \left( \wedge _{i}>t\right) =\Pr \left[ \min \left\{ T_{i,1},\cdots
,T_{i,l}\right\} >t\right] \\ 
\text{ } \\ 
=\Pr \left( T_{i,1}>t\right) \cdots \Pr \left( T_{i,l}>t\right) \\ 
\text{ } \\ 
=\Pr \left( T>t\right) ^{l}=\bar{F}\left( t\right) ^{l}%
\end{array}%
\right.  \label{A20}
\end{equation}%
($-\infty <t<\infty $). Eq. (\ref{203}) and Eq. (\ref{A20}) imply that%
\begin{equation}
\left. 
\begin{array}{l}
\Pr \left( \tilde{\wedge}_{i}>t\right) =\Pr \left[ l\cdot \left( \wedge
_{i}-t_{\ast }\right) >t\right] \\ 
\text{ } \\ 
=\Pr \left( \wedge _{i}>t_{\ast }+\frac{t}{l}\right) =\bar{F}\left( t_{\ast
}+\frac{t}{l}\right) ^{l}%
\end{array}%
\right.  \label{A21}
\end{equation}%
($-\infty <t<\infty $). Differentiating Eq. (\ref{A21}) with respect to the
variable $t$ implies that the density function of the scaled random variable 
$\tilde{\wedge}_{i}$ is given by%
\begin{equation}
-\frac{d}{dt}\Pr \left( \tilde{\wedge}_{i}>t\right) =\bar{F}\left( t_{\ast }+%
\frac{t}{l}\right) ^{l}\cdot \bar{h}\left( t_{\ast }+\frac{t}{l}\right)
\label{A22}
\end{equation}%
($-\infty <t<\infty $), where $\bar{h}\left( t\right) =f\left( t\right) /%
\bar{F}\left( t\right) $. In what follows we use the shorthand notation $%
\bar{\epsilon}=$ $\bar{h}\left( t_{\ast }\right) $. Note that the two basic
requirements $0<F\left( t_{\ast }\right) <1$ and $0<f\left( t_{\ast }\right)
<\infty $ imply that: $0<\bar{\epsilon}<\infty $.

Now, apply proposition \ref{P} to the following setting: $n=c$, $\theta =l$,
and $X_{i}=\tilde{\wedge}_{i}$ ($i=1,\cdots ,c$). Eq. (\ref{A22}) implies
that%
\begin{equation}
f_{\theta }\left( x\right) =\underset{\kappa _{\theta }}{\underbrace{\bar{F}%
\left( t_{\ast }\right) ^{\theta }}}\cdot \underset{g_{\theta }\left(
x\right) }{\underbrace{\left[ \frac{\bar{F}\left( t_{\ast }+\frac{x}{\theta }%
\right) }{\bar{F}\left( t_{\ast }\right) }\right] ^{\theta }\cdot \bar{h}%
\left( t_{\ast }+\frac{x}{\theta }\right) }}  \label{A23}
\end{equation}%
($-\infty <x<\infty $). Note that%
\begin{equation}
\left. 
\begin{array}{l}
\left[ \frac{\bar{F}\left( t_{\ast }+\frac{x}{\theta }\right) }{\bar{F}%
\left( t_{\ast }\right) }\right] ^{\theta }=\left[ \frac{\bar{F}\left(
t_{\ast }\right) -f\left( t_{\ast }\right) \frac{x}{\theta }+o\left( \frac{1%
}{\theta }\right) }{\bar{F}\left( t_{\ast }\right) }\right] ^{\theta } \\ 
\text{ } \\ 
=\left[ 1-\frac{\bar{\epsilon}x}{\theta }+o\left( \frac{1}{\theta }\right) %
\right] ^{\theta }\underset{\theta \rightarrow \infty }{\longrightarrow }%
\exp \left( -\bar{\epsilon}x\right)%
\end{array}%
\right.  \label{A24}
\end{equation}%
($-\infty <x<\infty $). Eqs. (\ref{A23}) and (\ref{A24}) imply that%
\begin{equation}
\lim_{\theta \rightarrow \infty }g_{\theta }\left( x\right) =g\left(
x\right) :=\bar{\epsilon}\exp \left( -\bar{\epsilon}x\right)  \label{A25}
\end{equation}%
($-\infty <x<\infty $). Also, the asymptotic geometric coupling of Eq. (\ref%
{204}) implies that the asymptotic coupling of Eq. (\ref{A11}) holds with $%
\kappa =1$. Hence, the result of proposition \ref{P} holds with the
intensity function 
\begin{equation}
\lambda \left( x\right) =\bar{\epsilon}\exp \left( -\bar{\epsilon}x\right)
\label{A26}
\end{equation}%
($-\infty <x<\infty $). This proves proposition \ref{P1}.

\subsection{\label{A2}Proof of proposition \protect\ref{P2}}

Set $\mathcal{P}$ to be a Poisson process, over the real line, with
intensity function $\lambda \left( x\right) =\bar{\epsilon}\exp \left( -\bar{%
\epsilon}x\right) $ ($-\infty <x<\infty $) and exponent $\bar{\epsilon}%
=f\left( t_{\ast }\right) /\bar{F}\left( t_{\ast }\right) $. Consider the
number of points $N\left( t\right) $ of the Poisson process $\mathcal{P}$
that reside above a real threshold $t$. The Poisson-process statistics imply
that the number $N\left( t\right) $ is a Poisson-distributed random variable

with mean%
\begin{equation}
\left. 
\begin{array}{l}
\mathbf{E}\left[ N\left( t\right) \right] =\int_{t}^{\infty }\lambda \left(
x\right) dx \\ 
\text{ } \\ 
=\int_{t}^{\infty }\bar{\epsilon}\exp \left( -\bar{\epsilon}x\right) dx=\exp
\left( -\bar{\epsilon}t\right) \text{ .}%
\end{array}%
\right.  \label{A40}
\end{equation}

Now, consider the \emph{maximal point} $M$ of the Poisson process $\mathcal{P%
}$. This maximal point is no larger than the threshold $t$ if and only if no
points of the Poisson process $\mathcal{P}$ reside above this threshold: $%
\left\{ M\leq t\right\} \Leftrightarrow \left\{ N\left( t\right) =0\right\} $%
. Hence, as $N\left( t\right) $ is a Poisson-distributed random variable
with mean $\mathbf{E}\left[ N\left( t\right) \right] $, Eq. (\ref{A40})
implies that the distribution function of the maximal point $M$ is given by%
\begin{equation}
\Pr \left( M\leq t\right) =\exp \left[ -\exp \left( -\bar{\epsilon}t\right) %
\right]  \label{A41}
\end{equation}%
($-\infty <t<\infty $). The distribution function of Eq. (\ref{A41})
characterizes the Gumbel law. A `standard' Gumbel-distributed random
variable $\mathcal{G}$ is governed by the distribution function of Eq. (\ref%
{303}): $\Pr \left( \mathcal{G}\leq t\right) =\exp \left[ -\exp \left(
-t\right) \right] $ ($-\infty <$ $t<\infty $). Eqs. (\ref{A41}) and (\ref%
{303}) imply that the maximal point $M$ admits the following Gumbel
representation in law:%
\begin{equation}
M=\bar{\eta}\cdot \mathcal{G}\text{ ,}  \label{A43}
\end{equation}%
where 
\begin{equation}
\bar{\eta}=\frac{1}{\bar{\epsilon}}=\frac{\bar{F}\left( t_{\ast }\right) }{%
f\left( t_{\ast }\right) }\text{ .}  \label{A44}
\end{equation}

Proposition \ref{P1} established that the ensemble $\left\{ \tilde{\wedge}%
_{1},\cdots ,\tilde{\wedge}_{c}\right\} $ converges in law -- in the limit
of Eq. (\ref{204}) -- to the Poisson process $\mathcal{P}$. Consequently,
the maximum $\tilde{\wedge}_{\max }$ of the ensemble $\left\{ \tilde{\wedge}%
_{1},\cdots ,\tilde{\wedge}_{c}\right\} $ converges in law -- in the limit
of Eq. (\ref{204}) -- to the maximal point $M$ of the Poisson process $%
\mathcal{P}$. Hence, Eq. (\ref{A43}) proves proposition \ref{P2}.

\subsection{\label{A3}Proof of proposition \protect\ref{P3}}

For the random variable $\vee _{j}=\max \left\{ T_{1,j},\cdots
,T_{c,j}\right\} $ we have%
\begin{equation}
\left. 
\begin{array}{l}
\Pr \left( \vee _{j}\leq t\right) =\Pr \left[ \max \left\{ T_{1,j},\cdots
,T_{c,j}\right\} \leq t\right] \\ 
\text{ } \\ 
=\Pr \left( T_{1,j}\leq t\right) \cdots \Pr \left( T_{c,j}\leq t\right) \\ 
\text{ } \\ 
=\Pr \left( T\leq t\right) ^{c}=F\left( t\right) ^{c}%
\end{array}%
\right.  \label{A30}
\end{equation}%
($-\infty <t<\infty $). In turn, for the scaled random variable $\tilde{\vee}%
_{j}=c\cdot \left( \vee _{j}-t_{\ast }\right) $ Eq. (\ref{A30}) implies that%
\begin{equation}
\left. 
\begin{array}{l}
\Pr \left( \tilde{\vee}_{j}\leq t\right) =\Pr \left[ c\cdot \left( \vee
_{j}-t_{\ast }\right) \leq t\right] \\ 
\text{ } \\ 
=\Pr \left( \vee _{j}\leq t_{\ast }+\frac{t}{c}\right) =F\left( t_{\ast }+%
\frac{t}{c}\right) ^{c}%
\end{array}%
\right.  \label{A31}
\end{equation}%
($-\infty <t<\infty $). Differentiating Eq. (\ref{A31}) with respect to the
variable $t$ implies that the density function of the scaled random variable 
$\tilde{\vee}_{j}$ is given by%
\begin{equation}
\frac{d}{dt}\Pr \left( \tilde{\vee}_{j}\leq t\right) =F\left( t_{\ast }+%
\frac{t}{c}\right) ^{c}\cdot h\left( t_{\ast }+\frac{t}{c}\right)
\label{A32}
\end{equation}%
($-\infty <t<\infty $), where $h\left( t\right) =f\left( t\right) /F\left(
t\right) $. In what follows we use the shorthand notation $\epsilon =$ $%
h\left( t_{\ast }\right) $. Note that the two basic requirements $0<F\left(
t_{\ast }\right) <1$ and $0<f\left( t_{\ast }\right) <\infty $ imply that: $%
0<\epsilon <\infty $.

Now, apply proposition \ref{P} to the following setting: $n=l$, $\theta =c$,
and $X_{i}=\tilde{\vee}_{j}$ ($j=1,\cdots ,l$). Eq. (\ref{A32}) implies that%
\begin{equation}
f_{\theta }\left( x\right) =\underset{\kappa _{\theta }}{\underbrace{F\left(
t_{\ast }\right) ^{\theta }}}\cdot \underset{g_{\theta }\left( x\right) }{%
\underbrace{\left[ \frac{F\left( t_{\ast }+\frac{x}{\theta }\right) }{%
F\left( t_{\ast }\right) }\right] ^{\theta }\cdot h\left( t_{\ast }+\frac{x}{%
\theta }\right) }}  \label{A33}
\end{equation}%
($-\infty <x<\infty $). Note that%
\begin{equation}
\left. 
\begin{array}{l}
\left[ \frac{F\left( t_{\ast }+\frac{x}{\theta }\right) }{F\left( t_{\ast
}\right) }\right] ^{\theta }=\left[ \frac{F\left( t_{\ast }\right) +f\left(
t_{\ast }\right) \frac{x}{\theta }+o\left( \frac{1}{\theta }\right) }{%
F\left( t_{\ast }\right) }\right] ^{\theta } \\ 
\text{ } \\ 
=\left[ 1+\frac{\epsilon x}{\theta }+o\left( \frac{1}{\theta }\right) \right]
^{\theta }\underset{\theta \rightarrow \infty }{\longrightarrow }\exp \left(
\epsilon x\right)%
\end{array}%
\right.  \label{A34}
\end{equation}%
($-\infty <x<\infty $). Eqs. (\ref{A33}) and (\ref{A34}) imply that%
\begin{equation}
\lim_{\theta \rightarrow \infty }g_{\theta }\left( x\right) =g\left(
x\right) :=\epsilon \exp \left( \epsilon x\right)  \label{A35}
\end{equation}%
($-\infty <x<\infty $). Also, the asymptotic geometric coupling of Eq. (\ref%
{404}) implies that the asymptotic coupling of Eq. (\ref{A11}) holds with $%
\kappa =1$. Hence, the result of proposition \ref{P} holds with the
intensity function 
\begin{equation}
\lambda \left( x\right) =\epsilon \exp \left( \epsilon x\right)  \label{A36}
\end{equation}%
($-\infty <x<\infty $). This proves proposition \ref{P3}.

\subsection{\label{A4}Proof of proposition \protect\ref{P4}}

Set $\mathcal{P}$ to be a Poisson process, over the real line, with
intensity function $\lambda \left( x\right) =\epsilon \exp \left( \epsilon
x\right) $ ($-\infty <x<\infty $) and exponent $\epsilon =f\left( t_{\ast
}\right) /F\left( t_{\ast }\right) $. Consider the number of points $N\left(
t\right) $ of the Poisson process $\mathcal{P}$ that reside up to a real
threshold $t$. The Poisson-process statistics imply that the number $N\left(
t\right) $ is a Poisson-distributed random variable with mean%
\begin{equation}
\left. 
\begin{array}{l}
\mathbf{E}\left[ N\left( t\right) \right] =\int_{-\infty }^{t}\lambda \left(
x\right) dx \\ 
\text{ } \\ 
=\int_{-\infty }^{t}\epsilon \exp \left( \epsilon x\right) dx=\exp \left(
\epsilon t\right) \text{ .}%
\end{array}%
\right.  \label{A50}
\end{equation}

Now, consider the \emph{minimal point} $M$ of the Poisson process $\mathcal{P%
}$. This minimal point is larger than the threshold $t$ if and only if no
points of the Poisson process $\mathcal{P}$ reside up to this threshold: $%
\left\{ M>t\right\} \Leftrightarrow \left\{ N\left( t\right) =0\right\} $.
Hence, as $N\left( t\right) $ is a Poisson-distributed random variable with
mean $\mathbf{E}\left[ N\left( t\right) \right] $, Eq. (\ref{A50}) implies
that the survival function of the minimal point $M$ is given by%
\begin{equation}
\Pr \left( M>t\right) =\exp \left[ -\exp \left( \epsilon t\right) \right]
\label{A51}
\end{equation}%
($-\infty <t<\infty $). A `standard' Gumbel-distributed random variable $%
\mathcal{G}$ is governed by the distribution function of Eq. (\ref{303}): $%
\Pr \left( \mathcal{G}\leq t\right) =\exp \left[ -\exp \left( -t\right) %
\right] $ ($-\infty <$ $t<\infty $). Eqs. (\ref{A51}) and (\ref{303}) imply
that the minimal point $M$ admits the following Gumbel representation in law:%
\begin{equation}
M=-\eta \cdot \mathcal{G}\text{ ,}  \label{A53}
\end{equation}%
where 
\begin{equation}
\eta =\frac{1}{\epsilon }=\frac{F\left( t_{\ast }\right) }{f\left( t_{\ast
}\right) }\text{ .}  \label{A54}
\end{equation}

Proposition \ref{P3} established that the ensemble that the ensemble $%
\left\{ \tilde{\vee}_{1},\cdots ,\tilde{\vee}_{l}\right\} $ converges in law
-- in the limit of Eq. (\ref{404}) -- to the Poisson process $\mathcal{P}$.
Consequently, the minimum $\tilde{\vee}_{\min }=\min \left\{ \tilde{\vee}%
_{1},\cdots ,\tilde{\vee}_{l}\right\} $ of the ensemble $\left\{ \tilde{\vee}%
_{1},\cdots ,\tilde{\vee}_{l}\right\} $ converges in law -- in the limit of
Eq. (\ref{404}) -- to the minimal point $M$ of the Poisson process $\mathcal{%
P}$. Hence, Eq. (\ref{A53}) proves proposition \ref{P4}.

\newpage

\subsection{\label{A5}MATLAB\ code for Figure 1}

{\fontfamily{courier}\selectfont

\noindent \% This function computes the scaled MaxMin/eta\_bar 
\\
\\
\noindent \% N specifies the number of random matrices to be generated
\\
\noindent N=10\textasciicircum{5};
\\
\\
\noindent \% MaxMin will hold the N Max-Min values that will be computed
\\
\noindent MaxMin=zeros(1,N);
\\
\\
\noindent \% pd specifies the distribution of the random matrix entries 
\\
pd = makedist('Exponential','mu',1);
\\
\\
\noindent \% CDF\_t specifies the value of the cumulative distribution function at the anchor point 
\\
CDF\_t=1/5;
\\
\\
\noindent \% This computes the anchor point t by inverting cumulative distribution function 
\\
t=icdf(pd,CDF\_t);
\\
\\
\noindent\% l sets the number of links
\\
l=70;
\\
\\
\noindent\% c sets the number of chains via geometric coupling
\\
c=floor((1-CDF\_t)\textasciicircum(-l));
\\
\\
\noindent\% This for-loop generates the random matrices and computes the MaxMin 
\\
\noindent for k=1:N
\\
\indent    M=random(pd,c,l);
    \\
\indent   MaxMin(k)=max(min(M'));
    \\
\noindent end
\\
\\
\noindent\% This computes the coefficient eta\_bar
\\
eta\_bar=(1-CDF\_t)/pdf(pd,t);
\\
\\
\noindent\% This computes the scaled MaxMin/eta\_bar
\\
MaxMin=(MaxMin-t)*l/eta\_bar;
\\
\\}

\newpage

\begin{center}
{\Large Table 1}

\bigskip\ 

\begin{tabular}{||l||l||l||}
\hline\hline
& $%
\begin{array}{c}
\text{ } \\ 
\text{\textbf{Proposition \ref{P1}}} \\ 
\text{ }%
\end{array}%
$ & $%
\begin{array}{c}
\text{ } \\ 
\text{\textbf{Proposition }\ref{P3}} \\ 
\text{ }%
\end{array}%
$ \\ \hline\hline
$%
\begin{array}{c}
\text{ } \\ 
\text{1. Quantity} \\ 
\text{ }%
\end{array}%
$ & $\wedge _{i}=\min \left\{ T_{i,1},\cdots ,T_{i,l}\right\} $ & $\vee
_{j}=\max \left\{ T_{1,j},\cdots ,T_{c,j}\right\} $ \\ \hline\hline
$%
\begin{array}{c}
\text{ } \\ 
\text{2. Scaling} \\ 
\text{ }%
\end{array}%
$ & $\tilde{\wedge}_{i}=l\cdot \left( \wedge _{i}-t_{\ast }\right) $ & $%
\tilde{\vee}_{j}=c\cdot \left( \vee _{j}-t_{\ast }\right) $ \\ \hline\hline
$%
\begin{array}{c}
\text{ } \\ 
\text{3. Ensemble} \\ 
\text{ }%
\end{array}%
$ & $\left\{ \tilde{\wedge}_{1},\cdots ,\tilde{\wedge}_{c}\right\} $ & $%
\left\{ \tilde{\vee}_{1},\cdots ,\tilde{\vee}_{l}\right\} $ \\ \hline\hline
$%
\begin{array}{c}
\text{ } \\ 
\text{4. Coupling} \\ 
\text{ }%
\end{array}%
$ & $\lim_{c\rightarrow \infty ,l\rightarrow \infty }c\cdot \bar{F}\left(
t_{\ast }\right) ^{l}=1$ & $\lim_{l\rightarrow \infty ,c\rightarrow \infty
}l\cdot F\left( t_{\ast }\right) ^{c}=1$ \\ \hline\hline
$%
\begin{array}{c}
\text{ } \\ 
\text{5. Intensity} \\ 
\text{ }%
\end{array}%
$ & $\lambda \left( x\right) =\bar{\epsilon}\exp \left( -\bar{\epsilon}%
x\right) $ & $\lambda \left( x\right) =\epsilon \exp \left( \epsilon
x\right) $ \\ \hline\hline
$%
\begin{array}{c}
\text{ } \\ 
\text{6. Exponent} \\ 
\text{ }%
\end{array}%
$ & $\bar{\epsilon}=f\left( t_{\ast }\right) /\bar{F}\left( t_{\ast }\right) 
$ & $\epsilon =f\left( t_{\ast }\right) /F\left( t_{\ast }\right) $ \\ 
\hline\hline
\end{tabular}

\bigskip\ 
\end{center}

\textbf{Table 1}: Summary of proposition \ref{P1} and proposition \ref{P3}.
Rows 1-3 summarize the underlying settings: the quantities under
consideration, their affine scalings, and the ensembles under consideration.
Rows 4-6 summarize the Poisson-process limit-law results: the required
asymptotic geometric couplings of $c$ and $l$, the intensity functions of
the limiting Poisson processes (to which the ensembles converge in law), and
their exponents.

\newpage

\begin{center}
{\Large Table 2}

\bigskip\ 

\begin{tabular}{||l||l||l||}
\hline\hline
& $%
\begin{array}{c}
\text{ } \\ 
\text{\textbf{Proposition }\ref{P2}} \\ 
\text{ }%
\end{array}%
$ & $%
\begin{array}{c}
\text{ } \\ 
\text{\textbf{Proposition }\ref{P4}} \\ 
\text{ }%
\end{array}%
$ \\ \hline\hline
$%
\begin{array}{c}
\text{ } \\ 
\text{1. Quantity} \\ 
\text{ }%
\end{array}%
$ & $\wedge _{\max }=\max \left\{ \wedge _{1},\cdots ,\wedge _{c}\right\} $
& $\vee _{\min }=\min \left\{ \vee _{1},\cdots ,\vee _{l}\right\} $ \\ 
\hline\hline
$%
\begin{array}{c}
\text{ } \\ 
\text{2. Scaling} \\ 
\text{ }%
\end{array}%
$ & $\tilde{\wedge}_{\max }=l\cdot \left( \ \wedge _{\max }-t_{\ast }\right) 
$ & $\tilde{\vee}_{\min }=c\cdot \left( \vee _{\min }-t_{\ast }\right) $ \\ 
\hline\hline
$%
\begin{array}{c}
\text{ } \\ 
\text{3. Coupling} \\ 
\text{ }%
\end{array}%
$ & $\lim_{c\rightarrow \infty ,l\rightarrow \infty }c\cdot \bar{F}\left(
t_{\ast }\right) ^{l}=1$ & $\lim_{l\rightarrow \infty ,c\rightarrow \infty
}l\cdot F\left( t_{\ast }\right) ^{c}=1$ \\ \hline\hline
$%
\begin{array}{c}
\text{ } \\ 
\text{4. Limit} \\ 
\text{ }%
\end{array}%
$ & $\lim_{c\rightarrow \infty ,l\rightarrow \infty }$ $\tilde{\wedge}_{\max
}=\bar{\eta}\cdot \mathcal{G}$ & $\lim_{l\rightarrow \infty ,c\rightarrow
\infty }\tilde{\vee}_{\min }=-\eta \cdot \mathcal{G}$ \\ \hline\hline
$%
\begin{array}{c}
\text{ } \\ 
\text{5. Coefficient} \\ 
\text{ }%
\end{array}%
$ & $\bar{\eta}=\bar{F}\left( t_{\ast }\right) /f\left( t_{\ast }\right) $ & 
$\eta =F\left( t_{\ast }\right) /f\left( t_{\ast }\right) $ \\ \hline\hline
\end{tabular}

\bigskip\ 
\end{center}

\textbf{Table 2}: Summary of proposition \ref{P2} and proposition \ref{P4}.
Rows 1-2 summarize the quantities under consideration and their affine
scalings. Rows 3-5 summarize the Gumbel limit-law results: the required
asymptotic geometric couplings of $c$ and $l$, the limiting Gumbel random
variables (the convergences being in law), and the coefficients of the
limiting Gumbel random variables. The term $\mathcal{G}$ appearing in row 4
is the `standard' Gumbel random variable of Eq. (\ref{303}).

\newpage

\begin{center}
{\Large Table 3}

\bigskip\ 

\begin{tabular}{||l||l||l||}
\hline\hline
& $%
\begin{array}{c}
\text{ } \\ 
\text{\textbf{Max-Min}} \\ 
\text{ }%
\end{array}%
$ & $%
\begin{array}{c}
\text{ } \\ 
\text{\textbf{Min-Max}} \\ 
\text{ }%
\end{array}%
$ \\ \hline\hline
$%
\begin{array}{c}
\text{ } \\ 
\text{1. Asymptotics} \\ 
\text{ }%
\end{array}%
$ & $l\gg1$ \ \& $\ c\cdot \bar{F}\left( t_{\ast }\right) ^{l}\simeq $ $1$ & $%
c\gg1$ \ \& $\ l\cdot F\left( t_{\ast }\right) ^{c}\simeq 1$ \\ \hline\hline
$%
\begin{array}{c}
\text{ } \\ 
\text{2. Approximation} \\ 
\text{ }%
\end{array}%
$ & $\mathcal{G}_{\ast }:=t_{\ast }+\frac{\bar{\eta}}{l}\cdot \mathcal{G}$ & 
$\mathcal{G}_{\ast }:=t_{\ast }-\frac{\eta }{c}\cdot \mathcal{G}$ \\ 
\hline\hline
$%
\begin{array}{c}
\text{ } \\ 
\text{3. Coefficient} \\ 
\text{ }%
\end{array}%
$ & $\bar{\eta}=\bar{F}\left( t_{\ast }\right) /f\left( t_{\ast }\right) $ & 
$\eta =F\left( t_{\ast }\right) /f\left( t_{\ast }\right) $ \\ \hline\hline
$%
\begin{array}{c}
\text{ } \\ 
\text{4. Mode} \\ 
\text{ }%
\end{array}%
$ & $t_{\ast }$ & $t_{\ast }$ \\ \hline\hline
$%
\begin{array}{c}
\text{ } \\ 
\text{5. Median} \\ 
\text{ }%
\end{array}%
$ & $t_{\ast }-\ln \left[ \ln \left( 2\right) \right] \bar{\eta}\cdot \frac{1%
}{l}$ & $t_{\ast }+\ln \left[ \ln \left( 2\right) \right] \eta \cdot \frac{1%
}{c}$ \\ \hline\hline
$%
\begin{array}{c}
\text{ } \\ 
\text{6. Mean} \\ 
\text{ }%
\end{array}%
$ & $t_{\ast }+\gamma \bar{\eta}\cdot \frac{1}{l}$ & $t_{\ast }-\gamma \eta
\cdot \frac{1}{c}$ \\ \hline\hline
$%
\begin{array}{c}
\text{ } \\ 
\text{7. SD} \\ 
\text{ }%
\end{array}%
$ & $\frac{\pi \bar{\eta}}{\sqrt{6}}\cdot \frac{1}{l}$ & $\frac{\pi \eta }{%
\sqrt{6}}\cdot \frac{1}{c}$ \\ \hline\hline
\end{tabular}
\end{center}

\bigskip\ 

\textbf{Table 3}: Summary of the Max-Min and the Min-Max Gumbel asymptotic
approximations of Eqs. (\ref{304}) and (\ref{406}). Rows 1-3 summarize the
approximations: the required asymptotics, the resulting approximations, and
the coefficients of the magnitudes of the approximations' stochastic parts.
The term $\mathcal{G}$ appearing in row 2 is the `standard' Gumbel random
variable of Eq. (\ref{303}). Rows 4-7 summarize the approximations' key
statistical features: modes, medians, means, and standard deviations (SD).
The term $\gamma $ appearing in row 6 is the Euler-Mascheroni constant: $%
\gamma =0.577\cdots $.

\newpage

\begin{center}
{\Large Table 4}

\bigskip\ 

\begin{tabular}{||l||l||l||l||}
\hline\hline
& $%
\begin{array}{c}
\text{ } \\ 
\text{\textbf{Max-Min}} \\ 
\text{ }%
\end{array}%
$ & $%
\begin{array}{c}
\text{ } \\ 
\text{\textbf{Min-Max}} \\ 
\text{ }%
\end{array}%
$ & $%
\begin{array}{c}
\text{ } \\ 
\text{\textbf{Normal}} \\ 
\text{ }%
\end{array}%
$ \\ \hline\hline
$%
\begin{array}{c}
\text{ } \\ 
\text{1. Approximation} \\ 
\text{ }%
\end{array}%
$ & $\mathcal{G}_{\ast }:=t_{\ast }+\frac{\bar{\eta}}{l}\cdot \mathcal{G}$ & 
$\mathcal{G}_{\ast }:=t_{\ast }-\frac{\eta }{c}\cdot \mathcal{G}$ & $%
\mathcal{N}_{\ast }:=\mu +\frac{\sigma }{\sqrt{n}}\cdot \mathcal{N}$ \\ 
\hline\hline
$%
\begin{array}{c}
\text{ } \\ 
\text{2. Deterministic} \\ 
\text{ }%
\end{array}%
$ & $t_{\ast }$ & $t_{\ast }$ & $\mu =\int_{-\infty }^{\infty }xf\left(
x\right) dx$ \\ \hline\hline
$%
\begin{array}{c}
\text{ } \\ 
\text{3. Coefficient} \\ 
\text{ }%
\end{array}%
$ & $\bar{\eta}=\bar{F}\left( t_{\ast }\right) /f\left( t_{\ast }\right) $ & 
$\eta =F\left( t_{\ast }\right) /f\left( t_{\ast }\right) $ & $\sigma =\sqrt{%
\int_{-\infty }^{\infty }\left( x-\mu \right) ^{2}f\left( x\right) dx}$ \\ 
\hline\hline
$%
\begin{array}{c}
\text{ } \\ 
\text{4. Order} \\ 
\text{ }%
\end{array}%
$ & $O\left( 1/l\right) $ & $O\left( 1/c\right) $ & $O\left( 1/\sqrt{n}%
\right) $ \\ \hline\hline
\end{tabular}

\end{center}

\bigskip\ 

\textbf{Table 4}: Summary of the similarities between the Max-Min and the Min-Max Gumbel asymptotic approximations of Eqs. (\ref{304}) and (\ref{406}) (for a preset reference time $t_{\ast }$), and the Normal asymptotic approximation of Eq. (\ref{612}). In row 1: the term $\mathcal{G}$ is the `standard' Gumbel
random variable of Eq. (\ref{303}), and the term $\mathcal{N}$ is a
`standard' Normal random variable (i.e. with zero mean and with unit
variance). Rows 2-3 summarize the approximations' structures: their
deterministic parts, the coefficients of the magnitudes of their stochastic
parts, and the orders of the magnitudes of their stochastic parts. For the
Normal column: $f\left( x\right) $ ($-\infty <x<\infty $) denotes the
density function of the input $X$, and $n$ is the size of the ensemble $%
\left\{ X_{1},\cdots ,X_{n}\right\} $ (see section \ref{6} for the details).

\newpage

\bigskip

\begin{figure}[t]

\includegraphics[width=7.5cm]{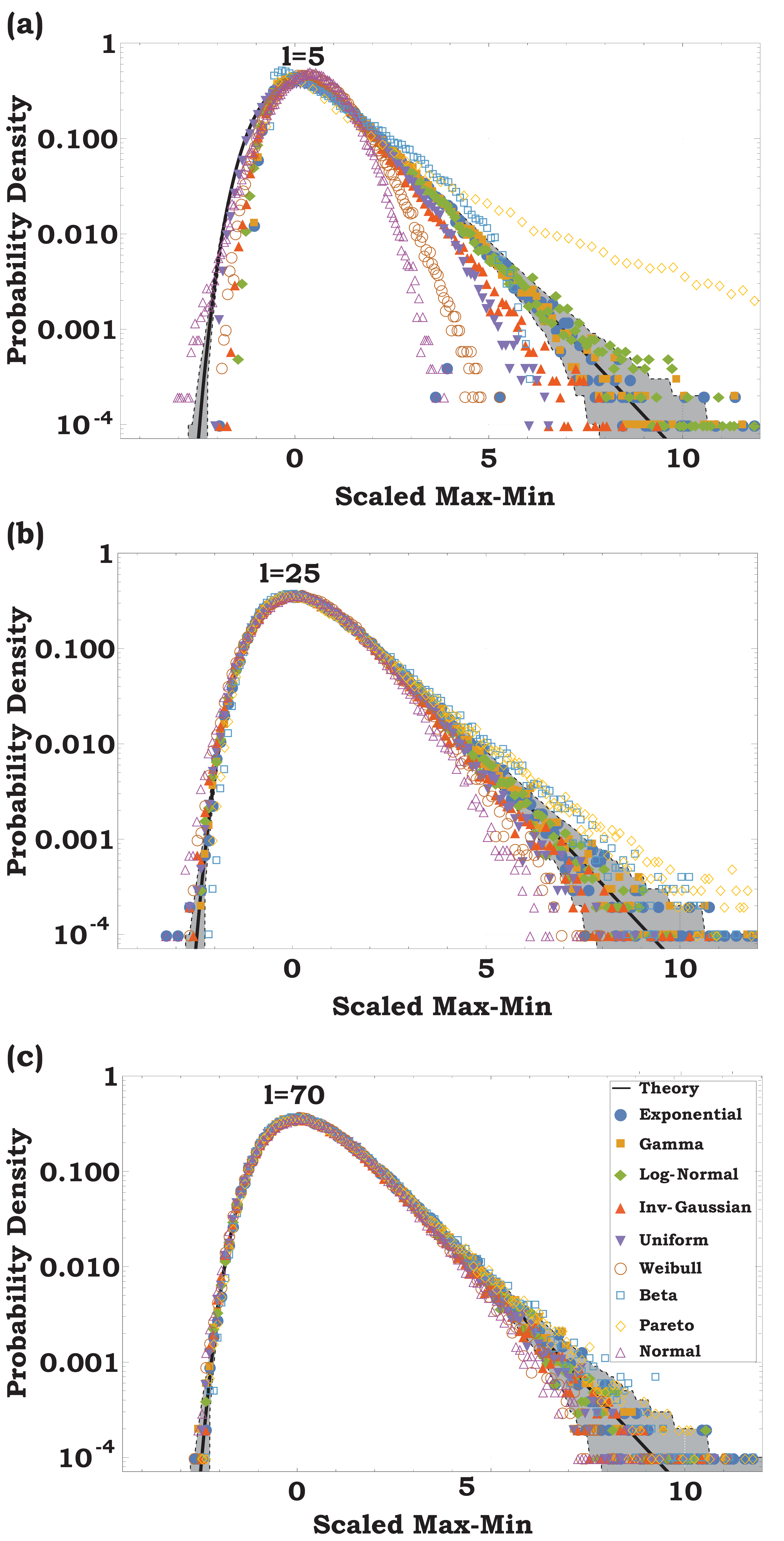}

\bigskip\ 

\justify \textbf{Figure 1}: Numerical simulations demonstrating Proposition \ref{P2} -- the convergence of the scaled Max-Min $\tilde{\wedge}_{\max }/\bar{\eta}$, in law, to the `standard' Gumbel random variable $\mathcal{G}$ (see section \ref{3} for the details). $10^{5}$ random matrices are simulated, with $l=5,25,70$ links and $c\simeq 1.25^{l}$ chains. The colored symbols depict the simulated data points of the scaled Max-Min. The solid black line depicts the density function of the `standard' Gumbel random variable $\mathcal{G}$ (with its $95\%$ confidence interval shaded in grey). Nine different distributions of the generic failure time $T$ are considered: Exponential, Gamma, Log-Normal, Inverse-Gauss, Uniform, Weibull, Beta,
Pareto, and Normal. As the number of links grows from $l=5$ (top) to $l=25$ (middle) and to $l=70$ (bottom): the convergence of the simulated data to the `standard' Gumbel density function is evident.
\label{fig1}
\end{figure}

\end{document}